# Preliminary simulation of temperature evolution in comminution processes in ball mills

F. Nurdiana[a], Muhandis[b], A. S. Wismogroho[c], N.T. Rochman[c], L.T. Handoko[a,b]

[a]Group for Nuclear and Particle Physics, Department of Physics, University of Indonesia, Kampus Depok, Depok 16424, Indonesia

[b]Group for Theoretical and Computational Physics, Research Center for Physics, Indonesian Institute of Sciences, Kompleks Puspiptek Serpong, Tangerang 15314, Indonesia
E-mail : laksana.tri.handoko@lipi.go.id

[c]Group for Nanomaterial, Research Center for Physics, Indonesian Institute of Sciences, Kompleks Puspiptek Serpong, Tangerang 15314, Indonesia

**ABSTRACT**

A model for comminution processes in ball mills is presented. The model provides an empirical approach using the physically realistic modelization of the ball mill system, but also deploys the statistical method to extract relevant thermodynamics observables. Those physical observables are investigated by treating the whole system as a statistical macroscopic ensemble.

## 1. INTRODUCTION

The comminution processes in recent years attract the attention among scientists and engineers due to the increasing demand of ultrafine powders for nanotechnology applications in many areas. The demand then requires the improvement of comminution equipments like ball mills, roller mills and so on. Unfortunately, the development of such comminution equipments always contains a lot of uncertainties due to a wide range of unknown parameters. These, in fact, lead to significant statistical errors. In order to overcome such problems, several models have been developed to quantitatively describe comminution process in various types of mills [1,2,3,4,5].

On the other hand, mathematical modeling and simulation may provide prior information and constraint to the unknown parameter ranges which should be useful to develop more optimized experimental strategy in comminution processes. However, in most cases of mathematical models, the physical observables like grain-size etc are extracted from a set of equation of motions (EOM). Such EOM's are considered to govern as complete as possible the dynamics of the system, from the mechanical motions to the evolution of grain-size distribution. This approach is obviously suffered from the nonlinearities of the equations under consideration, and then the requirement of high computational power to solve them numerically. This fact often discourages a quantitative and deterministic approach for the simulation of such system. These nonlinear effects like chaotic behavior of the sphere motions within the mill encourages some works modeling the system using semi-empirical approaches [6]. However most of semi-empirical models require a large number of experimental data based on prior observations [7], or measured variables obtained from simulation results by other authors [8,9].

In this paper we deploy a novel model and approach developed recently which is combining the deterministic approach for milling bodies motion, and the statistical approach to relate them with considerable macroscopic physical observables [10].

## 2. THE UNDERLYING MODEL

The whole system is modeled empirically using hamiltonian method. It is represented by a total hamiltonian describing all aspects of dynamics in the ball mills. It is further used to calculate the partition function and extracting the relevant thermodynamics observables.

The dynamics of each 'matter' in the system, i.e. balls and powders inside the vial, is described by a hamiltonian $H_m(r,t)$. The index m denotes the powder (p) or ball (b) and $r = (x,y,z)$. The hamiltonian contains some terms representing all relevant interactions working on the matters inside the system [10],

$$H = H_0 + V_{int} + V_{ext} \qquad (1)$$

with v denotes the vial, while $H_0$ is the free matter hamiltonian and $V$'s are the internal and external potentials.

The internal potential contains all considerable interactions among matters, i.e. among powder-powderm, powder-ball, ball-ball, powder-vial and ball-vial. The details of all potentials can be found in [10].

Using statistical mechanis, the hamiltonian is related to the partition function,

$$Z \propto \exp[-\beta H], \qquad (2)$$

with $\beta = 1/k_B T$ for temperature $T$ independent $H$. Further, the partition function is used to extract some statistical observables, for instance Helmholz free energy,

$$F_H = -1/\beta \ln Z, \qquad (3)$$

and pressure,

$$P = -F_H/V. \qquad (4)$$

Please note that the above equations hold for classical and general system. For a canonical system like our case, it is represented in a continue form in term of space displacement $x$ and momentum $p$ respectively [10].

Considering the normalized partition function, finally we found the pressure to be,

$$P = 1 - F\beta \ln^{-1}(2m\pi/\beta), \qquad (5)$$

where the auxiliary function $F$ represents all geometrical contribution of vial and its motions.

## 3. RESULT AND DISCUSSION

Making use of Eq. (5) we can draw general behavior of pressure in term of system temperature. Note that this result is independent from the geometrical form and motions of vial. The behavios is depicted in Fig. 1 as a function of system temperature $T$.

From the figure, it is clear that the pressure depends highly on the system temperture. Its rough behavior is same for any ball mills, but the shapes are determined completely by the function $F$ characterizing the geometrical structures and motions in the inertial system. Therefore Eq. (5) provides a general behavior for temperature-dependent pressure in the model, while the geometrical structure and motion of vial is absorbed in the function $F$. Its physically meaningfull region is depicted in Fig. 1 for any temperature along the horizontal axis.

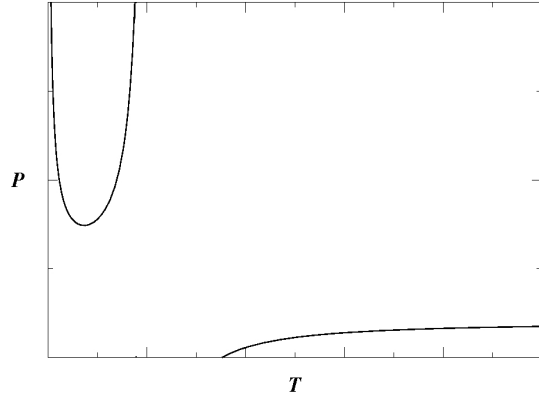

Fig. 1 The normalized temperature-dependent pressure in the system of any ball mills.

From the figure, the physically meaningful region is between $0 < T < 1/(2m \pi k_B)$, and $T > T_{th}$ where the threshold temperature is always greater than $1/(2m \pi k_B)$ respectively.

## 7. CONCLUSION

We have considered a model for communution processes in ball mills using hamiltonian approach. This approach has a great advantage on extracting the physical observables directly from the hamiltonian without a need to solve a set of equation of motions of the whole system which always leads to unsolvable non-linear problems.

More complete numerical calculation is required to take into account the contribution of geometrical shapes and motions of vial represented by the function $F$.

## ACKNOWLEDGMENT

The work is supported by the Riset Kompetitif LIPI in fiscal year 2009. Muhandis thanks the Group for Theoretical and Computational Physics LIPI for warm hospitality during the work.

## REFERENCES

[1] B. K. Mishra, R. K. Rajamani, "The discrete element method for the simulation of ball mills", *Applied Mathematical Modeling* 16 (1992) 598-604.
[2] B. K. Mishra, R. K. Rajamani, "Simulation of charge motion in ball mills", *International Journal of Mineral Processing* 40 (1994) 171-186.
[3] B. K. Mishra, "Charge dynamics in planetary mill", *Kona Powder Particle* 13 (1995) 151-158.
[4] B. K. Mishra, C. V. R. Murty, "On the determination of contact parameters for the

realistic {DEM} simulations of ball mills", *Powder Technology* 115 (2001) 290--297.
[5] T. Poschel, C. Saluena, "Scaling properties of granular materials", *Physical Review* E64 (2001) 011308.
[6] G.Manai, F.Delogu, M.Rustici, "Onset of chaotic dynamics in a ball mill : atractor merging and crisis induced intermittency", *Chaos* 12 (2002) 601-609.
[7] R.M. Davis, B.McDermott, C.C. Koch, "Mechanical alloying of brittle materials", *Metallurgical Transactions* A19 (1988) 2867.
[8] D.Maurice, T.H. Courtney, "The physics of mechanical alloying : a first report", *Metallurgical Transactions* A21 (1990) 289-302.
[9] D.Maurice, T.H. Courtney, "Milling dynamics, Part II : dynamic of a spex mill in a one dimensional mill", *Metallurgical Transactions* A27 (1996) 1981.
[10] Muhandis, F. Nurdiana, A. S. Wismogroho, N. T. Rochman, L. T. Handoko, "Extracting physical observables using macroscopic ensemble in the spex-mixer/mill simulation ", *American Institute of Physics Proceeding* (2009) in press.